\documentclass[11pt, a4paper]{article}
\usepackage{fullpage}

\usepackage[utf8]{inputenc}
\usepackage{subcaption}
\usepackage{amsthm,amssymb}
\usepackage{amsmath}
\usepackage{xcolor}
\usepackage{tikz,ifthen}
\usetikzlibrary{arrows.meta, positioning,calc}
\usepackage{hyperref}
\hypersetup{
    colorlinks=true,       
    linkcolor=blue,        
    citecolor=red,         
    filecolor=magenta,     
    urlcolor=cyan,         
    linktocpage=true
}

\usepackage{comment}
\usepackage{mathtools}
\newcommand{\cB}{{\mathcal{B}}}
\newcommand{\cC}{{\mathcal{C}}}
\newcommand{\id}{{\mathrm{id}}}
\newcommand{\True}{{\mathsf{True}}}
\newcommand{\False}{{\mathsf{False}}}

\usepackage{array}
\usepackage{cleveref}
\usepackage{cite}
\usepackage{arydshln,booktabs}
\theoremstyle{plain}
\newtheorem{theorem}{Theorem}[section]
\newtheorem{lemma}[theorem]{Lemma}
\newtheorem{claim}[theorem]{Claim}
\newtheorem{corollary}[theorem]{Corollary}

\theoremstyle{definition}

\newtheorem{remark}[theorem]{Remark}

\def\NP{\mathsf{NP}}
\newcommand{\problemdef}[3]{
    \begin{center}
    \fbox {   \parbox[c]{0.85\textwidth}{
        \textsc{\large #1} 
        
         \textbf{Input:} #2 \\
         \textbf{Goal:} #3 
        }}
    \end{center}
}
\definecolor{orangeDark}{RGB}{204,76,2}
\def\final{0}  
\def\iflong{\iffalse}
\ifnum\final=0  
\newcommand{\mnote}[1]{{\color{purple}[{\tiny \textbf{Mirabel:} \bf #1}]\marginpar{\color{purple}*}}}
\newcommand{\csnote}[1]{{\color{green}[{\tiny \textbf{Csaba:} \bf #1}]\marginpar{\color{green}*}}}
\newcommand{\ynote}[1]{{\color{olive}[{\tiny \textbf{Yutaro:} \bf #1}]\marginpar{\color{olive}*}}}
\newcommand{\fnote}[1]{{\color{orange}[{\tiny \textbf{Florian:} \bf #1}]\marginpar{\color{orange}*}}}
\newcommand{\enote}[1]{{\color{cyan}[{\tiny \textbf{Eszti:} \bf #1}]\marginpar{\color{cyan}*}}}
\else 
\newcommand{\mnote}[1]{}
\newcommand{\csnote}[1]{}
\newcommand{\ynote}[1]{}
\newcommand{\fnote}[1]{}
\newcommand{\enote}[1]{}
\fi
\usepackage{titling}
\title{Odd and Even Harder Problems on Cycle-Factors}

\thanksmarkseries{alph}
\author{%
Florian Hörsch\thanks{CISPA Helmholtz Center for Information Security, Saarbrücken, Germany. Email: \texttt{florian.hoersch@cispa.de}}
\hspace{1em} \and
Csaba Kir\'aly\thanks{HUN-REN--ELTE Egerv\'ary Research Group on Combinatorial Optimization, Budapest, Hungary. Email: \texttt{csaba.kiraly@ttk.elte.hu}}
\hspace{1em} \and
Mirabel Mendoza-Cadena\thanks{Instituto de Ciencias de la Ingeniería, Universidad de O'Higgins, Rancagua, Chile. Email: \texttt{mirabel.mendoza@uoh.cl}} 
\and
Gyula Pap\protect\thanks{Department of Operations Research, ELTE Eötvös Loránd University, Budapest, Hungary. Emails: \texttt{gyula.pap@ttk.elte.hu, szeti97@gmail.com}}
\hspace{1.5em} \and
Eszter Szabó\protect\footnotemark[4]
\hspace{1.5em} \and
Yutaro Yamaguchi\thanks{Graduate School of Information Science and Technology, Osaka University, Osaka, Japan. Email: \texttt{yutaro.yamaguchi@ist.osaka-u.ac.jp}}
}

\date{\empty}

\begin{document}

\maketitle
\thispagestyle{empty}

\begin{abstract}
 For a graph (undirected or directed), a \emph{cycle-factor} is a collection of vertex-disjoint cycles covering the entire vertex set. Cycle-factors subject to parity constraints arise naturally in the study of structural graph theory and algorithmic complexity. In this work, we study four variants of the problem of finding a cycle-factor subject to the following parity constraints: (1) all cycles are odd, (2) all cycles are even, (3) at least one cycle is odd, and (4) at least one cycle is even. We show that all the variants of the problem are NP-complete both in undirected and directed graphs, even if each vertex is incident to at most three edges. We also prove that the first two variants are NP-complete even for planar directed graphs.

\end{abstract}

\newpage
\pagenumbering{roman}
\tableofcontents
\newpage
\pagenumbering{arabic}
\setcounter{page}{1}

\section{Introduction}
Congruency constraints have a well-established role in classical combinatorial optimization. Recent interest in congruency-constrained problems has been fueled by their deep connections with integer programming (e.g., \cite{artmann2020thesis}), and their relevance in classical problems. 
For example, the problem of minimum cuts in a given graph that satisfy certain parity and congruency constraints has been studied \cite{BARAHONA1987213,ngele,moor}. Further research focused on finding paths \cite{iwata2022finding,2024OddPath,kawase2020twoforbidpath,kobayashi2017finding,schlotter2025odd} and cycles \cite{THOMAS2023228,wollan2011} with congruency constraints. In general, congruency-constrained variants of well-studied combinatorial optimization problems often require a fundamentally different analysis, as these constraints can alter the nature of the problem to the extent that classical techniques no longer apply. This is exemplified by the \textsc{Exact Matching} problem introduced by Papadimitriou and Yannakakis in 1982~\cite{papadimitriou1982complexity}, for which the existence of a deterministic polynomial-time algorithm remains an open question.

We study parity-restricted cycle-factors in undirected and directed graphs.
Let $G$ be an undirected or directed graph with vertex set $V(G)$ and edge set $E(G)$ or arc set $ A(G)$, respectively; we use the term ``edge'' in the undirected sense and ``arc'' in the directed sense.
We denote by $uv$ an edge between $u$ and $v$ and an arc from $u$ to $v$; in the latter case, $u$ and $v$ are called the \emph{tail} and the \emph{head}, respectively. 
For our purpose of studying cycle-factors, we assume that at most two parallel edges exist between each pair of vertices in an undirected graph, and that no parallel arcs with the same head and tail exist in a directed graph; we do not allow any selfloops in both undirected and directed graphs. 

When $G$ is directed, we say that $P= (s = v_1, v_2, v_3, \dots, v_{k+1} = t)$ with $k \ge 0$ is an \emph{$(s,t)$-path} if
\begin{itemize}
    \item $ v_i v_{i+1} \in A(G) $ for every $ i \in \{1, 2, \dots, k\} $, and
    \item all vertices $v_i$ are pairwise distinct, where the only possible exception is $s = t$.
\end{itemize}
If $s = t$ and $k \ge 2$, we call $P$ a \emph{cycle}.
The \emph{length} of $P$ is $k$, and $P$ is called \emph{odd} (or \emph{even}) if $k$ is odd (or even, respectively).
We sometimes use the terms \emph{digraph}, \emph{dipath}, and \emph{dicycle} to emphasize that the directed case is considered.
When $G$ is undirected, $P$ is an \emph{$(s,t)$-path} if each edge in $E(G)$ can be oriented so that $P$ is a dipath in the resulting digraph.
Cycles, length, and parity are defined analogously.
In particular, when $s = t$ and $k = 2$, two parallel edges between $v_1$ and $v_2$ are needed for $P$ to be a cycle.
We denote by $V(P)$, $A(P)$, and $E(P)$ the sets of vertices, of arcs, and of edges, respectively, that form $P$.

For a graph (undirected or directed), a \emph{cycle-factor} is a collection of vertex-disjoint cycles covering the vertex set. 
A combinatorial problem for finding a cycle-factor of a (undirected or directed) graph is simply stated as follows:

\problemdef{Cycle-Factor}
    {A graph.}
    {Find a cycle-factor.}

Cycles with parity constraints naturally appear in both structural graph theory and algorithmic complexity.
We consider eight variants of \textsc{Cycle-Factor}, namely the following four variants in each of the undirected and directed cases.

\problemdef{$\forall$Odd Cycle-Factor}
    {A graph.}
    {Find a cycle-factor such that it contains only odd cycles.}
\problemdef{$\forall$Even Cycle-Factor}
    {A graph.}
    {Find a cycle-factor such that it contains only even cycles.}    
\problemdef{$\exists$Odd Cycle-Factor}
    {A graph.}
    {Find a cycle-factor such that it contains at least one odd cycle.}
\problemdef{$\exists$Even Cycle-Factor}
    {A graph.}
    {Find a cycle-factor such that it contains at least one even cycle.}

Note that all problems (decision versions) are clearly in $\NP$, and hence $\NP$-hardness implies $\NP$-completeness.

\subsection{Related Work}
\label{sec:related_results}
\paragraph{Cycle-factors.}
The existence of a cycle-factor in $2k$-regular undirected graphs was shown by Petersen~\cite{petersen1891theorie} for $k \ge 1$.
Belck~\cite{belck1950factor} and Gallai~\cite{gallai1950factorisation} were the first to characterize undirected graphs admitting a cycle-factor. 
Their proofs imply simple algorithms in which the problem is reduced to finding a perfect matching (see, e.g., \cite{schrijver2003combinatorial}).

It is folklore that a directed graph has a cycle-factor if and only if the undirected bipartite graph constructed as follows has a perfect matching: split each vertex $v$ into an in-copy $v^-$ and an out-copy $v^+$ and replace each arc $uv$ by an edge between $u^+$ and $v^-$.
Bang-Jensen, Guo, and Yeo~\cite{bangJensen2000complementary} established conditions characterizing when a directed graph admits a cycle-factor, as a consequence of their work on complementary dicycles, i.e., two vertex-disjoint dicycles that cover all the vertices.
See, e.g., \cite[Sec. 5.7, 6.10]{bangJensen2008BookDigraphs} for further results for specific directed graphs such as tournaments.

The problem \textsc{Cycle-Factor} may be considered on mixed graphs, which can contain both edges and arcs.
In Section~\ref{sec:mixed}, we discuss its natural formulation and the $\NP$-hardness based on the result of Bang-Jensen and Casselgren~\cite{bang2015restricted}\footnote[1]{The authors have noticed the previous work \cite{bang2014arc,bang2015restricted} after submitting the earlier version~\cite{horsch2025odd} of this paper to arXiv in October 2025.}.

The two parity variants \textsc{$\forall$Odd Cycle-Factor} and \textsc{$\forall$Even Cycle-Factor} in directed graphs were considered by Bang-Jensen and Bessy~\cite{bang2014arc}\footnotemark[1].
They showed that both problems are $\NP$-hard via a reduction from the $2$-linkage problem in directed graphs \cite{fortune1980directed}.
We present alternative reductions that can impose stronger restrictions on the input graph.
Very recently, one of the variants was used to derive the $\NP$-hardness of a computational problem arising from a game~\cite{horiyama2026computational}.

\paragraph{$C_k$-free 2-factors.} 
A cycle-factor of an undirected graph is also called a \emph{$2$-factor}.
A 2-factor is called \emph{$C_k$-free} if it does not contain cycles of length $k$ or less.
The complexity of finding a $C_k$-free 2-factor has also been well investigated.
For instance, if $k$ is at least half of the number of vertices, a $C_k$-free $2$-factor is exactly a Hamiltonian cycle; thus, the problem is $\NP$-hard. 
The problem is also $\NP$-hard for any fixed $k \ge 5$ in general~\cite{cornuejols1980matching}, and is polynomial-time solvable when $k = 3$ \cite{hartvigsen1984extensions,hartvigsen2024finding} or when $k = 4$ and the input graph is bipartite \cite{hartvigsen2006finding,babenko2012improved,pap2007combinatorial}; it is open when $k = 4$ and the input graph is general.
There exists an extensive literature on different scenarios (see, e.g., \cite{Babenko2010triangle,Takazawa2017decomposition,Takazawa2017Finding,boyd2013finding}).

\paragraph{Odd cycles in graphs.} 
It is easy to find an odd cycle in a graph (directed or undirected).
More precisely, this can be done by DFS in linear time.
Moreover, a shortest one can be found based on BFS in polynomial time; see, e.g., \cite{schrijver2003combinatorial}.

\paragraph{Even cycles in undirected graphs.}
It is not difficult to observe that every connected undirected graph with no even cycle is a \emph{cactus}, i.e., a graph in which every edge is contained in at most one cycle.
We can easily determine in polynomial time whether there exists an even cycle in a given undirected graph based on this characterization.
Arkin, Papadimitriou, and Yannakakis~\cite{arkin1991modularity} gave a linear-time algorithm for a more general congruency constraint.

\paragraph{Even cycles in directed graphs.}
Robertson, Seymour, and Thomas~\cite{robertson1999permanents}, and independently McCuaig~\cite{McCuaig2004}, developed polynomial-time algorithms for detecting even dicycles by using the connection with Pfaffian orientations in bipartite graphs~\cite{VY89}. 
Important applications have emerged since then.
Using the characterization of bipartite graphs that have a Pfaffian orientation, Guenin and Thomas~\cite{guenin2011packing} gave an excluded-minor characterization of directed graphs such that in any subgraph the maximum number of disjoint dicycles coincides with the minimum size of a transversal of dicycles (i.e., a vertex set whose removal makes the graph acyclic).
Gorsky, Kawarabayashi, Kreutzer, and Wiederrecht~\cite{gorsky2024packing} and Gorsky~\cite{gorsky2024thesis} recently proved relaxed versions of the so-called Erd\H{o}s--P\'{o}sa property for even dicycles, which established a number of further implications. 
Björklund, Husfeldt, and Kaski~\cite{bjorklund2024shortest} proposed a randomized polynomial-time algorithm for computing a shortest even dicycle by advanced algebraic methods.

\subsection{Our Results and Organization}
We show that all of the four variants are $\NP$-hard in the directed case even when each vertex has at most three incident arcs (Theorem~\ref{thm:directed}).
By observing that there exists a reduction from directed graphs to undirected graphs that preserves the number of even and odd cycles, we also establish the $\NP$-hardness results for the undirected case even when each vertex has at most three incident edges (Theorem~\ref{thm:undirected}).
Furthermore, for the two variants \textsc{$\forall$Odd Cycle-Factor} and \textsc{$\forall$Even Cycle-Factor}, we provide alternative proofs showing that the problems remain $\NP$-hard even when the input digraph is planar (Theorem~\ref{thm:directed_planar}).

The rest of the paper is organized as follows.
We prepare basic terminology and notation in \Cref{sec:preliminaries}.
We prove hardness for directed graphs in \Cref{sec:directed}, and extend these results to undirected graphs in \Cref{sec:undirected}.
We discuss the mixed graph case in Section~\ref{sec:mixed}.

\section{Preliminaries}
\label{sec:preliminaries}
For an integer $n\geq 0$, we use $[n]=\{ 1, 2, \dots, n\}$.

Let $G$ be an undirected graph. 
For $F \subseteq E(G)$, we denote by $V(F)$ the set of vertices that are incident to at least one edge of $F$.
We denote the \emph{degree} of a vertex $v$ (i.e., the number of edges incident to $v$) by $\deg_G(v)$ or simply by $\deg(v)$.
Given $F \subseteq E(G)$ and $v \in V(G)$, we use $\deg_F(v)$ for $\deg_H(v)$, where $H$ is defined by $V(H)=V(G)$ and $E(H)=F$.
Next, $F$ is a cycle-factor of $G$ if and only if $\deg_F(v) = 2$ for all $v \in V(G)$; thus, a cycle-factor of an undirected graph is often referred to as a \emph{$2$-factor}.
We say $G$ is \emph{cubic} if $\deg_G(v) = 3$ for all $v \in V(G)$ and \emph{subcubic} if $\deg_G(v) \le 3$ for all $v \in V(G)$.

We employ analogous definitions for directed graphs.
Let $D$ be a directed graph. 
For $F \subseteq A(D)$, we denote by $V(F)$ the set of vertices that are incident to at least one arc of $F$.
We denote the \emph{in-degree} of a vertex $v$ by $\deg_D^{-}(v) = |\{\, uv \in A(D) \mid u \in V(D) \,\}|$ and the \emph{out-degree} of $v$ by $\deg_D^{+}(v) = |\{\, vu \in A(D) \mid u \in V(D) \,\}|$, and analogously define $\deg_F^-(v)$ and $\deg_F^+(v)$ for $F \subseteq A(D)$.
Next, $F$ is a cycle-factor of $D$ if and only if $\deg_F^-(v) = \deg_F^+(v) = 1$ for all $v \in V(D)$; we refer to a cycle-factor of a directed graph as a \emph{$(1,1)$-factor}.
We say $D$ is \emph{cubic} if $\deg^-_D(v) + \deg^+_D(v) = 3$ for all $v \in V(D)$ and \emph{subcubic} if $\deg^-_D(v) + \deg^+_D(v) \le 3$ for all $v \in V(D)$.

For a $2$-factor $F$ and a $(1, 1)$-factor $\vec{F}$, we denote by $\cC(F)$ and $\cC(\vec{F})$ the sets of cycles in $F$ and in $\vec{F}$, respectively.

\section{Directed Problems}
\label{sec:directed}

Our goal in this section is to prove the following theorem.

\begin{theorem}\label{thm:directed}
    \textsc{$\forall$Odd $(1,1)$-Factor}, \textsc{$\forall$Even $(1,1)$-Factor}, \textsc{$\exists$Odd $(1,1)$-Factor}, and \textsc{$\exists$Even $(1,1)$-Factor} are $\NP$-hard even for subcubic digraphs.
\end{theorem} 

The proof of Theorem~\ref{thm:directed} is sketched as follows.
We start with \textsc{3-SAT}, which is the most fundamental $\NP$-complete problem (Karp~\cite{karp2010reducibility}), and reduce it to \textsc{Width-5 Permutation Program (W5PP)} based on Barrington's theorem~\cite{BARRINGTON1989150} in Section~\ref{sec:SAT_to_permutation}.
Next, we describe a reduction from \textsc{W5PP} to one of our problems, \textsc{$\exists$Even $(1,1)$-Factor}, in Section~\ref{sec:permutation_to_cycle_factor}.
Finally, we prove the correctness of the reduction and how to adjust it for the other three variants in Section~\ref{sec:proof_of_reduction}.

The first two variants \textsc{$\forall$Odd $(1,1)$-Factor} and \textsc{$\forall$Even $(1,1)$-Factor} remain $\NP$-hard even for planar digraphs (Theorem~\ref{thm:directed_planar}), which is shown in Section~\ref{sec:planar-proofs}.

\subsection{From $3$-SAT to Permutation Program}\label{sec:SAT_to_permutation}
We first briefly review \textsc{3-SAT}.
Let $X$ be a set of binary variables.
A \emph{clause} is a disjunction of \emph{literals}, $C = \ell_1 \lor \ell_2 \lor \dots \lor \ell_k$, where each literal $\ell_j$ is either a variable $x \in X$ or its negation $\neg x$.
A propositional formula on $X$ in \emph{conjunctive normal form (CNF)} can be written as $\varphi = C_1 \land C_2 \land \cdots \land C_m$, where each $C_i$ is a clause; we identify $\varphi$ with the set of clauses $\cC = \{\, C_j \mid j \in [m] \,\}$.
An \emph{assignment} is a function $\phi \colon X \to \{\True, \False\}$ that assigns a truth value to each variable. The assignment $\phi$ \emph{satisfies} $\varphi$ if $\varphi$ is evaluated to true under $\phi$.
For a further introduction, see, e.g., \cite{Cygan2015Parameterized,garey2002computers}. 
\problemdef{3-Satisfiability (3-SAT)}
    {A set $X$ of binary variables and a set $\mathcal{C}$ of 3-literal clauses over $X$.}
    {Decide whether there exists a satisfying assignment for $(X,\mathcal{C})$ or not.}

This problem is well-known as $\NP$-complete~\cite{karp2010reducibility}.
We reduce \textsc{3-SAT} to another problem, \textsc{Width-5 Permutation Program}.
To state the problem, we need several definitions; 
Figures~\ref{fig:example-gadgetPerm} and \ref{fig:whole-structure} may be helpful as examples.

Let $S_5$ be the set of permutations of $[5]$, in which the identity permutation is denoted by $\id$.
A permutation $\tau \in S_5$ is called a \emph{$k$-cycle} $(k \ge 2)$ if its graphical representation (regarding $\tau(i) = j$ as an arc from $i$ to $j$) contains exactly one nontrivial cycle of length $k$ and $5 - k$ selfloops.
Given a set $X$ of variables, 
a \emph{conditional permutation} is a tuple $(x, \sigma^0, \sigma^1)$ such that $x \in X$ and $\sigma^0, \sigma^1\in S_5$, which is evaluated as $\sigma^{\phi(x)}$ under an assignment $\phi$ to $X$ (where we regard $\True = 1$ and $\False = 0$).
For a sequence $\cB = ((x_1, \sigma_1^0, \sigma_1^1), \dots, (x_\ell, \sigma_\ell^0, \sigma_\ell^1))$ of conditional permutations and an assignment $\phi$ to $X$, we denote the permutation on $[5]$ obtained by its evaluation as
\[\cB(\phi) \coloneqq \sigma_\ell^{\phi(x_{\ell})} \circ \dots \circ \sigma_1^{\phi(x_{1})},\]
where the permutations act from right to left.

We are ready to state the problem \textsc{Width-5 Permutation Program}.

\problemdef{Width-5 Permutation Program (W5PP)}
    {A set $X$ of binary variables, a $5$-cycle $\tau \in S_5$, and a sequence $\cB$ of conditional permutations on $[5]$ such that $\cB(\phi) \in \{\id, \tau\}$ for any $\phi$.}
    {Decide whether there exists an assignment $\phi$ with $\cB(\phi) = \tau$ or not.}

We use Barrington's theorem~\cite{BARRINGTON1989150} (Theorem~\ref{thm:Barrington}) to see that \textsc{W5PP} is $\NP$-hard.
To state it precisely, we need further definitions.

A \emph{fan-in $2$} circuit $\cC$ is a layered representation of a propositional formula, which is processed from bottom to top, such that
\begin{itemize}
\item the bottom layer consists of the possible literals $x$ and $\neg x$ $(x \in X)$ with any number of outputs, which copy the truth value $\True$ or $\False$ determined by an assignment $\phi$;
\item each non-bottom layer contains any number of logical operations $\bullet \in \{\wedge, \vee\}$, each of which takes exactly two outputs $a, b$ from strictly lower layers as its input and outputs exactly one $a \bullet b$ (no copy is allowed);
\item the top layer has exactly one output, which corresponds to the evaluation of an assignment $\phi$, denoted by $\cC(\phi)$.
\end{itemize}
The number of layers is called the \emph{depth} of the circuit.
Note that any CNF of $m$ clauses each contains a constant number of literals can be represented as a fan-in $2$ circuit of depth $\mathrm{O}(\log m)$; after a constant number of layers for $\vee$ in the clauses, arrange $\wedge$ like a balanced tournament bracket.
Thus, the following theorem gives a polynomial-time reduction from \textsc{3-SAT} to \textsc{W5PP}. 

\begin{theorem}[Barrington~\cite{BARRINGTON1989150}]\label{thm:Barrington}
For any fan-in $2$ circuit $\cC$ of depth $d$ and any $5$-cycle $\tau \in S_5$, one can construct a sequence $\cB$ of conditional permutations on $[5]$ of length at most $4^d$, in time polynomial in $|\cB|$, such that
\[\cB(\phi) = \begin{cases} \id & (\cC(\phi) = \False),\\ \tau & (\cC(\phi) = \True). \end{cases}\]
\end{theorem}

\begin{corollary}
    \textsc{W5PP} is $\NP$-hard.
\end{corollary}

\subsection{From Permutation Program to $(1,1)$-Factor}\label{sec:permutation_to_cycle_factor}
In this section, we reduce \textsc{W5PP} to our problems in which each vertex has at most two incoming arcs and at most two outgoing arcs.
Here we describe a construction of a digraph for \textsc{$\exists$Even $(1,1)$-Factor}. Later, in Section~\ref{sec:proof_of_reduction}, we explain how the other three variants can be treated by adjusting the parities. 

Let $X$ be the set of variables and $\cB = (\, (x_k, \sigma_k^0, \sigma_k^1) \mid k \in [\ell] \,)$ be the sequence of conditional permutations.

\paragraph{Assumptions.}
For each $(x, \alpha, \beta) \coloneqq (x_k, \sigma_k^0, \sigma_k^1)$ in $\cB$, we may assume that either
\begin{itemize}
\item $\alpha = \beta$, or
\item $\alpha = \id$ and $\beta$ is a $\kappa$-cycle for some $\kappa \ge 2$.
\end{itemize}
Indeed, if $\alpha \neq \beta$, then $(x, \alpha, \beta)$ can be replaced by two conditional permutations, $(x, \id, \beta \circ \alpha^{-1})$ after $(x, \alpha, \alpha)$.
Moreover, $\beta \circ \alpha^{-1} \neq \id$ contains one or two nontrivial cycles in its graphical representation (as there are only five vertices).
If there is one, then $\beta \circ \alpha^{-1}$ is a $\kappa$-cycle for some $\kappa \ge 2$.
Otherwise, $(x, \id, \beta \circ \alpha^{-1})$ can be replaced by $(x, \id, \gamma_1)$ and $(x, \id, \gamma_2)$ such that each of $\gamma_1, \gamma_2 \in S_5$ is a $\kappa$-cycle for some $\kappa \ge 2$ and their nontrivial cycles are disjoint.

After the above normalization, we may also assume that $\ell$ is odd by adding to $\cB$ a trivial conditional permutation $(x, \id, \id)$ for some $x \in X$ if necessary.

\paragraph{Gadgets for Conditional Permutations.}
We prepare $5(\ell + 1)$ vertices $v_{k, j}$ $(k \in [\ell + 1],\ j \in [5])$ to emulate $\cB$. 
Let $(x, \alpha, \beta) \coloneqq (x_k, \sigma_k^0, \sigma_k^1)$ be the $k$-th conditional permutation in $\cB$.

When $\sigma_k^0 = \sigma_k^1 = \alpha$, we just add five arcs $e_{k, j}^0 = v_{k, j}v_{k+1, \alpha(j)}$ $(j \in [5])$. 
In this case, these arcs are unique outgoing arcs for $v_{k,j}$, which must be included in any $(1, 1)$-factor; this means choosing the permutation $\alpha$ regardless of $\phi(x_k)$.

Suppose that $\id = \sigma_k^0 \neq \sigma_k^1 = \beta$ and $\beta$ is a $\kappa$-cycle for some $\kappa \ge 2$; an illustration of this case is shown in Figure~\ref{fig:example-gadgetPerm}.
Let $C_\beta \subseteq [5]$ be the set of elements $j$ with $\beta(j) \neq j$.
We add five arcs $e_{k, j}^0 = v_{k, j}v_{k+1, j}$ $(j \in [5])$; including them in the $(1, 1)$-factor corresponds to choosing the permutation $\id$, i.e., setting $\phi(x_k) = \False$. 
In addition, we add $\kappa$ arcs $e_{k,j}^1 = v_{k, j}v_{k+1, \beta(j)}$ $(j \in C_\beta)$; including them in the $(1, 1)$-factor together with the $5 - \kappa$ arcs $e_{k, j}^0$ $(j \in [5] \setminus C_\beta)$ corresponds to choosing the permutation $\beta$, i.e., setting $\phi(x_k) = \True$. 
Furthermore, we pick an arbitrary element $j^* \in C_\beta$ and subdivide the arc $e_{k,j^*}^1$ with two intermediate vertices $a^k_{x}$ and $b^k_{x}$, i.e., replace $e_{k,j^*}^1$ with a dipath $(v_{k,j^*}, a^k_{x}, b^k_{x}, v_{k+1,\beta(j^*)})$ of length three.
In the remainder of the construction, we do not add any arcs leaving $a^k_{x}$ or entering $b^k_{x}$, which implies that the middle arc $e_k^{ab} = a^k_{x} b^k_{x}$ must be used in any $(1, 1)$-factor.
Moreover, as seen below (Claim~\ref{cl:necessary}), the remaining two arcs $e_{k,j^*}^{va} = v_{k,j^*}a^k_{x}$ and $e_{k,j^*}^{bv} = b^k_{x} v_{k+1,\beta(j^*)}$ are always in the same boat, i.e., for any $(1, 1)$-factor, either both arcs are used or neither is used. 

In each case, we define two arc sets $Q_k^0$ and $Q_k^1$, which correspond to the cases where $\phi(x_k) = \False$ and $\phi(x_k) = \True$, respectively, as follows:
\begin{align*}
    Q_k^0 &\coloneqq \{\, e_{k,j}^0 \mid j \in [5] \,\},\\
    Q_k^1 &\coloneqq \begin{cases}
        Q_k^0 & (\alpha = \beta),\\
        \{e_{k,j^*}^{va},\, e_{k,j^*}^{bv}\} \cup \{\, e_{k,j}^1 \mid j \in C_\beta \setminus \{j^*\} \,\} \cup \{\, e_{k,j}^0 \mid j \in [5] \setminus C_\beta \,\} & (\alpha \neq \beta).
    \end{cases}
\end{align*}
Also, let $Q_k \coloneqq Q_k^0 \cup Q_k^1$.

\begin{figure}[t]
\centering%
\begin{tikzpicture}[
  vtx/.style={circle,draw,fill=black,inner sep=1.5pt},
  lbl/.style={draw=none,fill=none,font=\scriptsize},
  txt/.style={draw=none,fill=none,font=\scriptsize,align=left},
   myArrow/.style={draw,line width = 1.5pt, -{Stealth[length=2mm]}}
]
  \foreach \j in {1,...,5} {
    \node[vtx] (uk\j)  at (0,{-0.5*\j}) {};
    \node[vtx] (uk1\j) at (2.3,{-0.5*\j}) {};
    \node[lbl,left=1pt of uk\j]   {$v_{k,\j}$};
    \node[lbl,right=1pt of uk1\j] {$v_{k+1,\j}$};
  }
  \node[vtx] (a) at (0.6,0.0) {};
  \node[vtx] (b) at (1.5,0.0) {};
  \node[lbl,above=1pt of a] {$a_x^k$};
  \node[lbl,above=1pt of b] {$b_x^k$};

  \draw[myArrow, thick] (uk1) -- (uk11);
  \draw[myArrow, thick] (uk2) -- (uk12);
  \draw[myArrow, thick] (uk3) -- (uk13);
  \draw[myArrow, thick] (uk4) -- (uk14);
  \draw[myArrow, thick] (uk5) -- (uk15);

  \draw[myArrow,orangeDark] (uk1) -- (a);
  \draw[myArrow,orangeDark] (a) -- (b);
  \draw[myArrow,orangeDark] (b) -- (uk12);
  \draw[myArrow,orangeDark] (uk2) -- (uk14);
  \draw[myArrow,orangeDark] (uk4) -- (uk15);
  \draw[myArrow,orangeDark] (uk5) -- (uk11);

    
\end{tikzpicture}
\caption{
Illustration of the gadget for a conditional permutation $(x, \id, \beta)$ with $\beta = \left(\begin{smallmatrix}1&2&3&4&5\\2&4&3&5&1\end{smallmatrix}\right)$, which fixes $3$ and
cyclically permutes the remaining four elements as $1 \to 2 \to 4 \to 5 \to 1$; thus $\beta$ is a $4$-cycle
($\kappa=4$) with $C_\beta = \{\, j \in [5] \mid \beta(j) \neq j \,\} = \{1,2,4,5\}$.
We fix $j^\ast = 1 \in C_\beta$ as the position to be subdivided.
The black arcs realize the $\id$-branch: $Q_k^0 = \{\, e_{k,j}^0 \mid j \in [5] \,\}$. 
The thick orange arcs (except for $e_k^{ab} = a_x^k b_x^k$) realize the $\beta$-branch: 
$Q_k^1 = \{e_{k,1}^{va}, e_{k,1}^{bv}\} \cup \{e_{k,2}^1, e_{k,4}^1, e_{k,5}^1\} \cup \{e_{k,3}^0\}$.}
\label{fig:example-gadgetPerm}
\end{figure}

\paragraph{Gadgets for Synchronization of Variables.} 
Let $x \in X$ be a variable, and $1 \le k_1 < k_2 < \dots < k_q \le \ell$ be the indices such that $x_{k_p} = x$ and $\sigma_{k_p}^0 \neq \sigma_{k_p}^1$ $(p \in [q])$.
When $q = 0$, the value of $x$ under any assignment $\phi$ does not affect $\cB(\phi)$; we then add no synchronization gadget for $x$ and set $R_x = R_x^0 = R_x^1 = \emptyset$.

Suppose that $q \ge 1$.
We add three additional vertices $a_x, b_x, c_x$ together with three arcs $f_x^{ac} = a_x c_x$, $f_x^{cb} = c_x b_x$, and $f_x^{ba} = b_x a_x$.
We also add $q + 1$ arcs $f_{x, 1}^{ba} = b_x a^{k_1}_x$, $f_{x, p + 1}^{ba} = b^{k_p}_x a^{k_{p+1}}_x$ $(p \in [q - 1])$, and from $f_{x, q + 1}^{ba} = b^{k_q}_x a_x$.
An illustration of this gadget is shown in Figure~\ref{fig:sync-gadget}.

Together with the arcs $e_{k_p}^{ab}$ $(p \in [q])$, these arcs without $f_x^{ba}$ form an odd cycle on $\{a_x, b_x, c_x\} \cup \{\, a_x^{k_p}, b_x^{k_p} \mid p \in [q] \,\}$.
As seen below (Claim~\ref{cl:necessary}), the case $\phi(x) = \False$ corresponds to including this cycle in a $(1, 1)$-factor, and when $\phi(x) = \True$, only $f_x^{ac}$, $f_x^{cb}$, $f_x^{ba}$, and $e_{k_p}^{ab}$ are used.
We define two corresponding arc sets $R_x^0$ and $R_x^1$, respectively, as follows:
\begin{align*}
    R_x^0 &\coloneqq \{f_x^{ac},\, f_x^{cb}\} \cup \{\, f_{x, p}^{ba} \mid p \in [q + 1] \,\} \cup \{\, e_{k_p}^{ab} \mid p \in [q] \,\},\\
    R_x^1 &\coloneqq \{f_x^{ac},\, f_x^{cb},\, f_x^{ba}\} \cup \{\, e_{k_p}^{ab} \mid p \in [q] \,\}.
\end{align*}
Also, let $R_x \coloneqq R_x^0 \cup R_x^1$.

\begin{figure}[t]
\centering
\begin{tikzpicture}[
  vtx/.style={circle,draw,fill=black,inner sep=1.5pt},
  lbl/.style={draw=none,fill=none,font=\scriptsize},
   myArrow/.style={draw,line width = 1.5pt, -{Stealth[length=2mm]}}
]
  \foreach \name/\x in {a/0, c/1, b/2, ak1/3, bk1/4, ak2/5, bk2/6, ak3/7, bk3/8} {
    \node[vtx] (\name) at (\x,0) {};
  }
  \node[lbl,above=2pt of a]   {$a_x$};
  \node[lbl,above=2pt of c]   {$c_x$};
  \node[lbl,above=2pt of b]   {$b_x$};
  \node[lbl,above=2pt of ak1] {$a_x^{k_1}$};
  \node[lbl,above=2pt of bk1] {$b_x^{k_1}$};
  \node[lbl,above=2pt of ak2] {$a_x^{k_2}$};
  \node[lbl,above=2pt of bk2] {$b_x^{k_2}$};
  \node[lbl,above=2pt of ak3] {$a_x^{k_3}$};
  \node[lbl,above=2pt of bk3] {$b_x^{k_3}$};

  \draw[myArrow] (a) -- (c) node[midway,below,lbl]{$f_x^{ac}$};
  \draw[myArrow] (c) -- (b) node[midway,below,lbl]{$f_x^{cb}$};
  \draw[myArrow] (ak1) -- (bk1) node[midway,below,lbl]{$e_{k_1}^{ab}$};
  \draw[myArrow] (ak2) -- (bk2) node[midway,below,lbl]{$e_{k_2}^{ab}$};
  \draw[myArrow] (ak3) -- (bk3) node[midway,below,lbl]{$e_{k_3}^{ab}$};

  \draw[myArrow,red,densely dotted] (b) -- (ak1) node[midway,below,lbl]{$f_{x,1}^{ba}$};
  \draw[myArrow,red,densely dotted] (bk1) -- (ak2) node[midway,below,lbl]{$f_{x,2}^{ba}$};
  \draw[myArrow,red,densely dotted] (bk2) -- (ak3) node[midway,below,lbl]{$f_{x,3}^{ba}$};
  \draw[myArrow,red,densely dotted] (bk3) to[bend right=40] node[midway,below,lbl]{$f_{x,4}^{ba}$} (a);

  \draw[myArrow,blue,dashed] (b) to[bend left=75] node[midway,below,yshift=2pt,lbl]{$f_x^{ba}$} (a);
\end{tikzpicture}
\caption{Synchronization gadget for a variable $x$ with $q=3$ occurrences $k_1<k_2<k_3$.
The thick black arcs must be included in any $(1, 1)$-factor.
The dotted red arcs realize the $\False$-branch: $R_x^0 = \{f_x^{ac}, f_x^{cb}\} \cup \{f_{x, 1}^{ba}, f_{x, 2}^{ba}, f_{x, 3}^{ba}, f_{x, 4}^{ba}\} \cup \{e_{k_1}^{ab}, e_{k_2}^{ab}, e_{k_3}^{ab}\}$.
The dashed blue arc realizes the $\True$-branch: $R_x^1 = \{f_x^{ac}, f_x^{cb}, f_x^{ba}\} \cup \{e_{k_1}^{ab}, e_{k_2}^{ab}, e_{k_3}^{ab}\}$.}
\label{fig:sync-gadget}
\end{figure}

\paragraph{Completing the Whole Structure.}
We finally complete the whole cycle structure by adding arcs from $v_{\ell+1, j}$ $(j \in [5])$ to some $v_{1, j'}$ so that this part corresponds to $\tau^{-1}$, as shown in Figure~\ref{fig:whole-structure}.
Recall that $\tau$ is an arbitrarily fixed $5$-cycle on $[5]$, and by symmetry we may assume that $\tau(j) \equiv j + 1 \pmod{5}$; then, $\tau^{-1}(j) \equiv j - 1 \pmod{5}$.
We add five arcs $e_{\ell+1, j} = v_{\ell + 1, j}v_{1, \tau^{-1}(j)}$ $(j \in [5])$.
Furthermore, we subdivide the arc $e_{\ell+1, 1}$ with one intermediate vertex $z$, i.e., replace $e_{\ell+1, 1}$ with a dipath $(v_{\ell+1,1}, z, v_{1, 5})$ of length two.
The arcs defined here are the unique outgoing or incoming arcs for some $v_{\ell + 1, j}$ or $v_{1, j}$, and hence all of them are included in any $(1, 1)$-factor.
Let $Q_{\ell + 1}$ be the set of those arcs.

Recall that we assume $\ell$ is odd.
Then, assuming that the other parts behave correctly as the evaluation of the conditional permutations in $\cB$ under an assignment $\phi$, a $(1,1)$-factor is always in the following situations:
\begin{itemize}
\item If $\cB(\phi) = \id$, there is a large odd cycle containing all $v_{k, j}$ $(k \in [\ell + 1],\ j \in [5])$.
\item If $\cB(\phi) = \tau$, there are five cycles each containing exactly one $v_{k, j}$ for each $k \in [\ell + 1]$, in which exactly one (through $z$) is odd and the others are even.
\end{itemize}

\begin{figure}[t!]
\centering
\begin{tikzpicture}[
  vtx/.style={circle,draw,fill=black,inner sep=1.5pt},
  lbl/.style={draw=none,fill=none,font=\scriptsize},
  myArrow/.style={draw,line width = 1.2pt, -{Stealth[length=2mm]}}
]
  \foreach \j in {1,...,5} {
    \node[vtx] (lk\j) at (0,-0.6*\j) {};
    \node[vtx] (l1\j) at (2.0,-0.6*\j) {};
    \node[lbl,left=2pt of lk\j]  {$v_{\ell+1,\j}$};
    \node[lbl,right=2pt of l1\j] {$v_{1,\j}$};
  }
  \node[vtx] (z) at (0.5,-0.6) {};
  \node[lbl,above=1pt of z] {$z$};

  \draw[myArrow] (lk1) -- (z);
  \draw[myArrow] (z) -- (l15);

  \draw[myArrow] (lk2) -- (l11);
  \draw[myArrow] (lk3) -- (l12);
  \draw[myArrow] (lk4) -- (l13);
  \draw[myArrow] (lk5) -- (l14);
\end{tikzpicture}
\caption{Closing arcs $Q_{\ell+1}$, corresponding to
$\tau^{-1} = \left(\begin{smallmatrix}1&2&3&4&5\\5&1&2&3&4\end{smallmatrix}\right)$,
the inverse of the fixed $5$-cycle
$\tau = \left(\begin{smallmatrix}1&2&3&4&5\\2&3&4&5&1\end{smallmatrix}\right)$.
Together with the $k$-th layers of Figure~\ref{fig:example-gadgetPerm} $(k \in [\ell])$ with synchronization of Figure~\ref{fig:sync-gadget}, closing
the loop this way turns the whole column structure into either one odd cycle when
$\cB(\phi)=\id$, or into four even cycles plus one odd cycle through $z$ when
$\cB(\phi)=\tau$.}
\label{fig:whole-structure}
\end{figure}

\subsection{$\NP$-hardness for Subcubic Graphs (Proof of Theorem~\ref{thm:directed})}\label{sec:proof_of_reduction}
We start the proof of Theorem~\ref{thm:directed}.
Let $D$ be the digraph constructed in Section~\ref{sec:permutation_to_cycle_factor}.

\subsubsection{Correctness of Reduction}
First, we observe several properties that bridge the assignments to $X$ and the $(1, 1)$-factors in $D$, which are intuitively easy from the construction.
Note that $A(D) = \left(\bigcup_{k = 1}^{\ell+1} Q_k\right) \cup \left(\bigcup_{x \in X} R_x\right)$ by definition.

\begin{claim}\label{cl:necessary}
    Let $F$ be a $(1, 1)$-factor of $D$.
    Then,
    \begin{enumerate}
        \item $F \supseteq Q_{\ell+1}$;
        \item $F \cap Q_k \in \{Q_k^0, Q_k^1\}$ for every $k \in [\ell]$;
        \item $F \cap R_x \in \{R_x^0, R_x^1\}$ for every $x \in X$.
    \end{enumerate}
\end{claim}

\begin{proof}
    1.~
    By construction, each arc in $Q_{\ell + 1}$ is a unique outgoing arc for some $v_{\ell+1, j}$ or a unique incoming arc for some $v_{1, j}$.
    Thus, these arcs must be included.

    2.~
    Let $(x, \alpha, \beta) \coloneqq (x_k, \sigma_k^0, \sigma_k^1) \in \cB$.
    If $\alpha = \beta$, then $Q_k = Q_k^0$ and each arc in $Q_k$ is a unique outgoing arc for some $v_{k,j}$.
    Thus, we conclude $F \supseteq Q_k$, equivalently, $F \cap Q_k = Q_k^0$.
    
    Suppose that $\id = \alpha \neq \beta$ and $F_k \coloneqq F \cap Q_k \neq Q_k^0$.
    We show $F_k = Q_k^1$.
    By construction, each $v_{k,j}$ $(j \in [5] \setminus C_\beta)$ has a unique outgoing arc $e_{k, j}^0 = v_{k,j}v_{k+1,j}$.
    Thus, $F_k \supseteq \{\, e_{k, j}^0 \mid j \in [5] \setminus C_\beta \,\}$.
    Since $F_k \neq Q_k^0$, there exists an arc $e_{k, j}^0 \in Q_k^0 \setminus F_k$ $(j \in C_\beta)$.
    Then, $F_k$ must contain an arc leaving $v_{k,j}$ which is distinct from $e_{k, j}^0$. Observe that there is exactly one such arc, namely $e_{k,j}^1$ if $j \neq j^*$ and $e_{k,j^*}^{va}$ if $j=j^*$.
    If $e_{k,j}^1 \in F_k$, then $e_{k,\beta(j)}^0 \not\in F_k$ as $\deg_{F_k}^-(v_{k+1,\beta(j)}) = 1$.
    Repeating this, we conclude that $e_{k,j^*}^{va} \in F_k$ since $\beta$ contains only one nontrivial cycle on $C_\beta \ni j^*$.
    Similarly, starting from the other side, we conclude that $e_{k, j^*}^{bv} \in F_k$.
    The uniqueness of the nontrivial cycle in $\beta$ implies that $F_k \supseteq \{\, e_{k,j}^1 \mid j \in C_\beta \setminus \{j^*\} \,\}$ and $F_k \cap \{\, e_{k,j}^0 \mid j \in C_\beta \,\} = \emptyset$, which completes the proof.
    
    3.~
    Suppose that $F_x \coloneqq F \cap R_x \neq R_x^0$.
    We show $F_x = R_x^1$.
    By construction, $f_x^{ac}$ and $f_x^{cb}$ are unique incoming and outgoing arcs for $c_x$, and hence $F_x \supseteq \{f_x^{ac}, f_x^{cb}\}$.
    In addition, each $e_{k_p}^{ab}$ $(p \in [q])$ is a unique outgoing arc for $a_x^{k_p}$, and hence $F_x \supseteq \{\, e_{k_p}^{ab} \mid p \in [q] \,\}$.
    Since $F_x \neq R_x^0$, we have $f_x^{ba} \in F_x$ or $f_{x,p}^{ba} \not\in F_x$ for some $p \in [q + 1]$.

    If $f_x^{ba} \in F_x$, then $f_{x,1}^{ba} \not\in F_x$ as $\deg_{F_x}^+(b_x) = 1$.
    In this case, $F$ must contain a unique alternative $e_{k_1}^{va}$ of an incoming arc for $a_x^{k_1}$, which implies by property 2 that $F$ contains $e_{k_1}^{bv}$ as well.
    Thus, $f_{x,2}^{ba} \not\in F_x$ as $\deg_{F_x}^+(b_x^{k_1}) = 1$.
    Repeating this, we conclude that $F_x \cap \{\, f_{x, p}^{ba} \mid p \in [q + 1] \,\} = \emptyset$, and hence $F_x = R_x^1$.

    It is easy to see that we conclude the same situation by starting with the latter case, i.e., if $f_{x,p}^{ba} \not\in F_x$ for some $p \in [q + 1]$.
    Thus, we complete the proof.
\end{proof}

\begin{claim}\label{cl:sufficient}
    Let $F \subseteq A(D)$ be an arc set satisfying the properties in Claim~\ref{cl:necessary}, i.e.,
    for some $\delta_k \in \{0, 1\}$ $(k \in [\ell])$ and $\epsilon_x \in \{0, 1\}$ $(x \in X)$,
    \begin{align}\label{eq:F}
        F = Q_{\ell + 1} \cup \left(\bigcup_{k=1}^\ell Q_k^{\delta_k}\right) \cup \left(\bigcup_{x \in X} R_x^{\epsilon_x}\right).
    \end{align}
    Then, $F$ is a $(1, 1)$-factor if and only if $\delta_k = \epsilon_{x_k}$ for every $k \in [\ell]$ with $\sigma_k^0 \neq \sigma_k^1$.
    Moreover, such a $(1, 1)$-factor $F$ contains an even cycle if and only if $\cB(\phi) = \tau$ for the assignment $\phi$ defined by $\phi(x) = \epsilon_x$ $(x \in X)$, where $\True = 1$ and $\False = 0$.
\end{claim}

\begin{proof}
The first equivalence is almost obvious from the proof of Claim~\ref{cl:necessary} for properties 2 and 3.
Indeed, if $F$ is a $(1, 1)$-factor and $F \cap Q_k = Q_k^0 \neq Q_k^1$, then $e_{k,j^*}^{va}$ and $e_{k,j^*}^{bv}$ are not in $F$, which implies $f_{x_k, p}^{ba} \in F \cap R_{x_k}$ for some $p$ and hence $F \cap R_{x_k} = R_{x_k}^0$.
If $F$ is a $(1, 1)$-factor and $F \cap Q_k = Q_k^1 \neq Q_k^0$, then $e_{k,j^*}^{va}$ and $e_{k,j^*}^{bv}$ are in $F$, which implies $f_{x_k, p}^{ba} \not\in F \cap R_{x_k}$ for some $p$ and hence $F \cap R_{x_k} = R_{x_k}^1$.
The converse direction is similar.

We show the second equivalence.
Let $F$ be a $(1, 1)$-factor in the form of \eqref{eq:F}, $\phi$ be an assignment defined by $\phi(x) = \epsilon_x$, and $\sigma \coloneqq \cB(\phi) \in \{\id, \tau\}$.
By construction and the first equivalence, $F$ contains five vertex-disjoint paths each from $v_{1,j}$ to $v_{\ell+1,\sigma(j)}$ $(j \in [5])$.
If $\sigma = \id$, they are connected into a single large cycle by $Q_{\ell+1}$, and otherwise (if $\sigma = \tau$), they are closed one-by-one by $Q_{\ell+1}$.
The large cycle and one of the five smaller cycles are odd, and the other four smaller cycles are even.
Since $F$ cannot have an even cycle in any other part by construction (as only two types of odd cycles in synchronization gadgets can appear in $F$), this implies the second equivalence.
\end{proof}

We now prove that the existence of the solutions is preserved.
For an assignment $\phi$ to $X$ with $\cB(\phi) = \tau$, by Claim~\ref{cl:sufficient}, we can construct a $(1, 1)$-factor $F$ in $D$ in which at least one cycle is even by \eqref{eq:F}: set $\epsilon_x = \phi(x)$ $(x \in X)$ and $\delta_k = \phi(x_k)$ $(k \in [\ell])$ with $\sigma_k^0 \neq \sigma_k^1$, and choose $\delta_k \in \{0, 1\}$ arbitrarily for $k \in [\ell]$ with $\sigma_k^0 = \sigma_k^1$.
Conversely, suppose that there exists a $(1, 1)$-factor $F$ in $D$ in which at least one cycle is even.
By Claim~\ref{cl:necessary}, $F$ is in the form of \eqref{eq:F}, and by Claim~\ref{cl:sufficient}, the assignment $\phi$ to $X$ defined by $\phi(x) = \epsilon_x$ $(x \in X)$ satisfies $\cB(\phi) = \tau$.
This completes the reduction for \textsc{$\exists$Even $(1,1)$-Factor}.

\paragraph{Adjusting Parities.}
The construction of $D$ given in Section~\ref{sec:permutation_to_cycle_factor} is for \textsc{$\exists$Even $(1,1)$-Factor}.
We can switch to \textsc{$\exists$Odd $(1,1)$-Factor} as follows:
\begin{itemize}
\item Split each $c_x$ into two copies $c_x^-$ and $c_x^+$ so that the dipath $(a_x, c_x, b_x)$ is replaced by $(a_x, c_x^-, c_x^+, b_x)$, which changes the parity of the cycles through $c_x$ from odd to even.
\item Change the assumption on $\ell$ from odd to even, which changes the parity of each of the cycles corresponding to $\cB(\phi)$ simultaneously.
\end{itemize}

We can also switch to \textsc{$\forall$Odd $(1,1)$-Factor} by modifying the arcs $e_{\ell+1, j}$ $(j \in [5])$ so that this part does not correspond to $\tau^{-1}$ but $\id$.
That is, we redefine five arcs $e_{\ell+1, j} \coloneqq v_{\ell + 1, j}v_{1, j}$ $(j \in [5])$, and subdivide the arc $e_{\ell+1, 1}$ with one intermediate vertex $z$, i.e., replace the arc with a dipath $(v_{\ell+1,1}, z, v_{1, 1})$ of length two.
Then, the situation changes as follows:
\begin{itemize}
\item If $\cB(\phi) = \id$, there are five cycles each containing exactly one $v_{k, j}$ for each $k \in [\ell + 1]$, in which exactly one (through $z$) is odd and the others are even.
\item If $\cB(\phi) = \tau$, there is a large odd cycle containing all $v_{k, j}$ $(k \in [\ell + 1],\ j \in [5])$.
\end{itemize}
Combining with the same modification above (from \textsc{$\exists$Even $(1,1)$-Factor} to \textsc{$\exists$Odd $(1,1)$-Factor}), we can switch to \textsc{$\forall$Even $(1,1)$-Factor} as well.

\subsubsection{Degree Restriction}
It is clear from the construction that $1 \le \deg_D^-(v) \le 2$ and $1 \le \deg_D^+(v) \le 2$ for every $v \in V(D)$.
Moreover, the input digraph $D$ can be transformed in polynomial time into an equivalent digraph $D'$ such that
\begin{itemize}
    \item $\min\{\deg_{D'}^-(v), \deg_{D'}^+(v)\} = 1$, and
    \item $\max_{v' \in V(D')}\max\{\deg_{D'}^-(v'), \deg_{D'}^+(v')\} = \max_{v \in V(D)}\max\{\deg_{D}^-(v), \deg_{D}^+(v)\}$.
\end{itemize}
Split each vertex $v \in V(D)$ into three copies $v^-, v', v^+$ with adding two arcs $v^-v'$ and $v'v^+$ and replace each original arc $uv$ with $u^+v^-$: 
this is almost the same transformation as the reduction from directed to undirected given in Section~\ref{sec:undirected} (cf.~Figure~\ref{fig:construction}).
For every $v \in V(D)$, we have $\deg_{D'}^+(v^-) = \deg_{D'}^-(v') = \deg_{D'}^+(v') = \deg_{D'}^-(v^+) = 1$, $\deg_{D'}^-(v^-) = \deg_D^-(v)$, and $\deg_{D'}^+(v^+) = \deg_D^+(v)$; thus, the desired conditions are indeed satisfied.
Moreover, since $\deg_{D'}^-(v') = \deg_{D'}^+(v') = 1$ implies that both arcs incident to $v'$ must be included in any $(1, 1)$-factor of $D'$, there exists a natural bijection between the $(1, 1)$-factors of $D$ and of $D'$, which preserves the parity of each dicycle (cf.~Claim~\ref{cl:undirected}).
This concludes that $D$ can be transformed to be subcubic without changing the answer.

\subsection{$\NP$-hardness of Two Variants for Planar Graphs}\label{sec:planar-proofs}
In this section, we prove the following theorem.

\begin{theorem}\label{thm:directed_planar}
\mbox{ }
    \begin{enumerate}
        \item \label{thmNPHardPlanarAllOdd} \textsc{$\forall$Odd $(1,1)$-Factor} is $\NP$-hard even for planar subcubic digraphs. 
        \item \label{thmNPHardPlanarAllEven} \textsc{$\forall$Even $(1,1)$-Factor} is $\NP$-hard even for planar digraphs.
    \end{enumerate}
\end{theorem} 

\subsubsection{$\forall$Odd Case.}
For the first variant \textsc{$\forall$Odd $(1,1)$-Factor}, we reduce the Hamiltonian cycle problem, which is known to be $\NP$-hard even for planar subcubic digraphs (Plesník~\cite{Plesnik1979}\footnote{It only states in the introduction that both in-degree and out-degree are at most two, but the digraph constructed in the reduction is indeed subcubic as remarked at the end of the proof.}).
Here, a path or cycle is \emph{Hamiltonian} if it is through all the vertices in the graph.

\problemdef{Planar Subcubic Hamiltonian Cycle}
    {A planar subcubic digraph $H$.}
    {Decide whether $H$ contains a Hamiltonian cycle or not.}

This problem is easily reduced to the following problem.
We may assume that every vertex has at least one incoming arc and at least one outgoing arc in $H$, since otherwise the answer is trivially No.
Pick one vertex $r \in V(H)$, split $r$ into two vertices $s$ and $t$ with an extra arc $ts$, and replace each arc $ur \in A(H)$ with $ut$ and $rv \in A(H)$ with $sv$.
The planarity and the degree restriction are clearly preserved.

\problemdef{Planar Subcubic Hamiltonian $(s,t)$-Path}
    {A planar subcubic digraph $H$ and $s, t \in V(H)$ with $ts \in A(H)$.}
    {Decide whether $H$ contains a Hamiltonian $(s,t)$-path or not.}

The following lemma proves the first statement of Theorem~\ref{thm:directed_planar}.
    
\begin{lemma}\label{lem:planar_all_odd}
    Given a planar subcubic digraph $H$ and $s, t \in V(H)$ with $ts \in A(H)$, one can construct in polynomial time a planar subcubic digraph $D$ such that $H$ has a Hamiltonian $(s, t)$-path if and only if $D$ has a $(1, 1)$-factor in which all the cycles are odd.
\end{lemma}

\begin{proof}
We construct a planar subcubic digraph $D$ in the following way. We let $V(D)$ consist of $V(H)$ and three extra vertices $x_1^{a}$, $x_2^{a}$, and $x_3^{a}$ for each $a \in A(H) \setminus \{ts\}$. Next, we let $A(D)$ consist of the sets of five arcs, $S_a=\{ux_1^a, x_1^ax_2^a,x_2^ax_3^a,x_3^ax_1^a,x_3^av\}$, for all $a=uv \in A(H) \setminus \{ts\}$ and the arc $ts$. This finishes the description of $D$. For an illustration, see Figure~\ref{fig:edge_gadget_all_odd}.
Since the new vertices $x_1^a,x_2^a,x_3^a$ introduced for an arc $a=uv$ are only adjacent to $u$ and $v$, the gadget $S_a$ can be drawn inside a thin disk following the curve of $a$ in a fixed planar embedding of $H$ without crossing any other arc of $A(D) \setminus \{ts\}$.
Also, since $ts \in A(H)$, the extra arc $ts \in A(D)$ can also be added without crossing any other arc of $A(D) \setminus \{ts\}$; hence $D$ is planar.
Moreover, $D$ is subcubic as $\deg_D^-(v) = \deg_H^-(v)$ and $\deg_D^+(v) = \deg_H^+(v)$ for every $v \in V(H)$ and all the extra vertices clearly have at most three incident arcs.
It is not difficult to see that $D$ can be constructed from $H$ in polynomial time. 

\begin{figure}[t!]
    \centering
    \begin{tikzpicture}[myNode/.style={circle, draw,  minimum size=2em},
    myEdge/.style={draw,line width = 1.5pt},
     myArrow/.style={draw,line width = 1pt, -{Stealth[length=3mm]}}]

        \node (x1) at (0,0) [myNode] {$u$};
        \node (x2) at (2,0) [myNode] {$x^{a}_1$};
        \node (x3) at (4,0) [myNode] {$x^{a}_2$};    
        \node (x4) at (6,0) [myNode] {$x^{a}_3$};
        \node (x5) at (8,0) [myNode] {$v$};
    
        \foreach \u\v in {x1/x2,x2/x3,x3/x4,x4/x5}{
            \draw[myArrow] (\u) to (\v); 
        }
        \draw[myArrow, bend left] (x4) to (x2);
    \end{tikzpicture}
    \caption{The gadget for an arc $a=uv \in A(H) \setminus \{ts\}$ in the reduction from \textsc{Planar Subcubic Hamiltonian $(s,t)$-Path} to \textsc{$\forall$Odd $(1,1)$-Factor}.}
    \label{fig:edge_gadget_all_odd}
\end{figure}

The following claim concludes the proof. 
\begin{claim}
    $(H, s, t)$ is a Yes-instance of \textsc{Planar Subcubic Hamiltonian $(s,t)$-Path} if and only if $D$ is a Yes-instance of \textsc{$\forall$Odd $(1, 1)$-Factor}.
\end{claim}

\begin{proof}
First, suppose that there exists a Hamiltonian $(s, t)$-path $P$ in $H$; note that $ts \not\in A(P)$.
We now let $F$ consist of the four arcs $ux_1^a,x_1^ax_2^a,x_2^ax_3^a,x_3^av$ for all $a=uv \in A(P)$, the arc $ts$, and the three arcs $x_1^ax_2^a,x_2^ax_3^a,x_3^ax_1^a$ for all $a \in A(H)\setminus (A(P) \cup \{ts\})$. It follows directly from the construction that $F$ is a $(1,1)$-factor of $D$. We still need to show that $C$ is odd for all $C \in \mathcal{C}(F)$. First, observe that for all $a \in A(H)\setminus (A(P) \cup \{ts\})$ the unique cycle  $C \in \mathcal{C}(F)$ with $V(C)=\{x_1^a,x_2^a,x_3^a\}$ contains three vertices and hence is clearly odd. The only directed cycle $C \in \mathcal{C}(F)$ that is not of that form satisfies $|V(C)|=|V(P)|+3|A(P)|=4|A(P)|+1$, which is also odd. Hence $D$ is a Yes-instance of \textsc{$\forall$Odd $(1,1)$-Factor}.

 Now suppose that there exists a $(1,1)$-factor $F$ of $D$ such that $C$ is odd for all $C\in \mathcal{C}(F)$.
 \begin{claim}
      \label{resiuiuoh}
     For every $a=uv \in A(H) \setminus \{ts\}$, exactly one of the following holds:
     \begin{itemize}
        \item $S_a \cap F =\{ux_1^a,x_1^ax_2^a,x_2^ax_3^a,x_3^av \}$,
        \item $S_a \cap F =\{x_1^ax_2^a,x_2^ax_3^a,x_3^ax_1^a \}$.
     \end{itemize}
 \end{claim}

\begin{proof}
 First observe that $\{x_1^ax_2^a,x_2^ax_3^a\}\subseteq F$ as $F$ is a $(1,1)$-factor of $D$ and these are the only arcs incident to $x_2^a$ in $D$. Next, if $x_3^ax_1^a\in F$, we obtain that $\{ux_1^a,x_3^av\}\cap F=\emptyset$ as $F$ is a $(1,1)$-factor of $D$. Finally, if $x_3^ax_1^a$ is not contained in $F$, we obtain that $\{ux_1^a,x_3^av\}\subseteq F$ as $F$ is a $(1,1)$-factor of $D$.
\end{proof}
 
 \begin{claim}
 \label{shadiasdia}
     Let $C\in \mathcal{C}(F)$ be a dicycle with $V(C)\cap V(H)\neq \emptyset$.
     Then, $ts \in A(C)$.
 \end{claim}

\begin{proof}
Suppose for the sake of a contradiction that there exists a directed cycle $C\in \mathcal{C}(F)$ with $V(C)\cap V(H)\neq \emptyset$ and such that $A(C)$ does not contain the arc $ts$. Let $A_0\subseteq A(H) \setminus \{ts\}$ consist of the arcs $a \in A(H) \setminus \{ts\}$ such that $A(C)\cap S_a \neq \emptyset$. As $C$ is a directed cycle and $V(C)\cap V(H)\neq \emptyset$, we obtain from Claim~\ref{resiuiuoh} that $|A(C)\cap S_a|=4$ for all $a \in A_0$. Hence, as $\{\, S_a \mid a \in A(H) \setminus \{ts\} \,\}$ is a partition of $A(D)\setminus \{ts\}$ and $A(C)$ does not contain $ts$, we obtain that $|V(C)|=4|A_0|$. In particular, we obtain that $C$ is even, contradicting the assumption on $F$.
\end{proof}

As $F$ is a $(1,1)$-factor, there exists a cycle $C$ in $\mathcal{C}(F)$ with $V(H)\cap V(C)\neq \emptyset$, which is unique as $ts \in A(C)$ by Claim~\ref{shadiasdia}. 
This means $V(H) \subseteq V(C)$.
Now let $P$ be the subgraph of $H$ such that $A(P)$ contains all $a=uv \in A(H) \setminus \{ts\}$ with $S_a\cap F=\{ux_1^a,x_1^ax_2^a,x_2^ax_3^a,x_3^av \}$. It follows by Claim~\ref{resiuiuoh} and the fact that $C$ is a dicycle in $D$ that $P$ is a dipath in $H$. 
As $V(P)=V(C)\cap V(H) = V(H)$, we obtain that $P$ is a Hamiltonian $(s, t)$-path in $H$, so $(H,s,t)$ is Yes-instance of \textsc{Hamiltonian $(s,t)$-path}.
\end{proof}

Thus, we complete the proof of Lemma~\ref{lem:planar_all_odd}.
\end{proof}

\subsubsection{$\forall$Even Case.}
We turn to the second variant \textsc{$\forall$Even $(1, 1)$-Factor}.
We reduce the following problem that was shown to be $\NP$-hard by Marx~\cite{marx2005}\footnote{In the list coloring problem, the number of colors (the size of the union of the color lists) is not necessarily bounded. Nevertheless, only three colors $1, 2, 3$ are used in the reduction in \cite{marx2005}, so we include this condition in the problem statement.}.
Here, for an undirected graph $H$ and an integer $k$, a mapping $\phi \colon E(H) \to [k]$ is a \emph{(proper) $k$-edge-coloring} if $\phi(e) \neq \phi(f)$ for every pair of distinct edges $e, f \in E(H)$ sharing at least one of their end vertices.
When a possible edge-color list $\Phi(e) \subseteq [k]$ for each edge $e \in E(H)$ is specified, it is called a \emph{list edge-coloring} if $\phi(e) \in \Phi(e)$ for every $e \in E(H)$.

    \problemdef{Planar Cubic List 3-Edge-Coloring}
        {A planar cubic bipartite undirected graph $H$ and edge-color lists $\Phi \colon E(H) \to 2^{[3]}$.}
        {Decide whether $H$ admits a list edge-coloring $\phi \colon E(H) \to [3]$ or not.}

The following lemma proves the second statement of Theorem~\ref{thm:directed_planar}.
    
\begin{lemma}\label{lem:planar_even_odd}
    Given a planar cubic bipartite undirected graph $H$ and edge-color lists $\Phi \colon E(H) \to 2^{[3]}$, one can construct in polynomial time a planar digraph $D$ such that $H$ admits a list edge-coloring $\phi \colon E(H) \to [3]$ if and only if $D$ has a $(1, 1)$-factor in which all the cycles are even.
\end{lemma}

\begin{proof}
    Let $(V^+, V^-)$ be the bipartition of the vertex set $V(H)$; as $H$ is regular, we have $|V^+| = |V^-|$.
    We construct a digraph $D$ by replacing each edge of $H$ by a gadget: the vertex set is $V(D)= V(H) \cup\{\, w'_e,w''_e \mid e \in E(H)  \,\}$ and the arc set $A(D)$ is the union of $A_{e} \coloneqq \{w'_e w''_e\} \cup \left(\bigcup_{i \in \Phi(e)}A_{e}^i\right)$ $(e \in E(H))$, where for each $e = uv \in E(H)$ with $u \in V^+$ and $v \in V^-$,
    \begin{align*}
        A_{e}^1 \coloneqq \{  uw'_{e}, w''_{e}v \},\quad A_{e}^2 \coloneqq \{  vw'_{e}, w''_{e}u \},\quad A_{e}^3 \coloneqq \{  w''_{e}w'_{e} \}.
    \end{align*}
    Note that the gadget contains a unique Hamiltonian path $ P_{uv} $ from $ u \in V^+ $ to $ v \in V^- $ if $1 \in \Phi(e)$ and $P_{vu}$ from $v \in V^-$ to $u \in V^+$ if $2 \in \Phi(e)$. 
    This finishes the description of $D$.
    An illustration of the gadget can be found in Figure~\ref{fig:edge_gadget_all_even}.
    Since the gadget for an edge $e = uv \in E(H)$ introduces only the two new vertices $w'_{e},w''_{e}$ and all arcs of $A_{e}$ have their end vertices in $u$, $v$, and these two vertices, the gadget can be drawn within a thin strip around the curve of $uv$ in a fixed planar embedding of $H$, replacing that edge without crossing any other part of the drawing; hence $D$ is planar.
    It is not difficult to see that $D$ can be computed from $H$ in polynomial time. 
    \begin{figure}[t!]
    \centering
    \begin{tikzpicture}[myNode/.style={circle, draw,  minimum size=2em},
    myEdge/.style={draw,line width = 1.5pt},
     myArrow/.style={draw,line width = 1pt, -{Stealth[length=3mm]}}]

        \node (u) at (0,0) [myNode] {$u$};
        \node (w1) at (2,0.8) [myNode] {$w'_{e}$};
        \node (w2) at (2,-0.8) [myNode] {$w''_{e}$};    
        \node (v) at (4,0) [myNode] {$v$};
    
        \foreach \u\v in {u/w1,w2/v}{
            \draw[myArrow, red, very thick, densely dotted] (\u) to (\v); 
        }
        \foreach \u\v in {v/w1,w2/u}{
            \draw[myArrow, blue, very thick, dashed] (\u) to (\v); 
        }
        \draw[myArrow, bend right, ultra thick] (w1) to (w2);
        \draw[myArrow, bend right, brown, thick] (w2) to (w1);
    \end{tikzpicture}
    \caption{The gadget for an edge $e = uv \in E(H)$ with $u \in V^+$ and $v \in V^-$ in the reduction from \textsc{Planar Cubic List 3-Edge-Coloring} to \textsc{$\forall$Even $(1, 1)$-Factor}. The thick black arc $w'_ew''_e$ always exists, the dotted red arcs correspond to color $1$, the dashed blue arcs correspond to color $2$, and the thin brown arc corresponds to color $3$.}
    \label{fig:edge_gadget_all_even}
\end{figure}

The following claim concludes the proof. 

\begin{claim}
    \label{vgudaiosian}
    $H$ is a Yes-instance of \textsc{Planar Cubic List 3-Edge-Coloring} if and only if $D$ is a Yes-instance of \textsc{$\forall$Even $(1, 1)$-Factor}.
\end{claim}
\begin{proof}
First, suppose that there exists a list 3-edge-coloring $\phi$ of $(H, \Phi)$. As $H$ is cubic, every vertex is incident to exactly three edges. As $\phi$ is a proper edge coloring, every edge incident to a fixed vertex has a different color. Every color defines a perfect matching, and so two colors define a collection of even cycles covering every vertex in $H$, that is, two colors define a $2$-factor of $H$ in which all the cycles are even.
In particular, by replacing each edge $e = uv \in E(H)$ with $u \in V^+$ and $v \in V^-$ by the corresponding path $P_{uv}$ of length $3$ in $D$ if $\phi(e) = 1$ or $P_{vu}$ if $\phi(e) = 2$, we obtain a collection of vertex-disjoint even dicycles in $D$ that covers all the vertices in $V(H)$.
By adding the dicycle of length $2$ of form $(w'_{e}, w''_{e}, w'_{e})$ for each edge $e$ of the remaining color $3$, we obtain a $(1, 1)$-factor of $D$ in which all the cycles are even. 

Now suppose that there exists a $(1, 1)$-factor $\vec{F}$ of $D$ in which all the cycles are even.
Since $H$ is cubic, each $u \in V(H)$ has exactly three neighbors $v_1, v_2, v_3 \in V(H)$ in $H$.
By the construction of $D$, for each $e \in E(H)$, a unique outgoing arc $w'_ew''_e$ for $w'_e$ must be included in $\vec{F}$.
In addition, since all the cycles in $\vec{F}$ are even, the cycle of form $(u, w'_e, w''_e, u)$ cannot be included.
Thus, we conclude that for each $u \in V(H)$, there exists an ordering $(i,j,k)$ of $[3]$ such that the following holds (where we employ the notation assuming $u \in V^+$ by symmetry):
\begin{itemize}
    \item $\vec{F} \cap A_{uv_i} = A(P_{uv_i}) = \{w'_{uv_i}w''_{uv_i}\} \cup A_{uv_i}^1$,
    \item $\vec{F} \cap A_{uv_j} = A(P_{v_ju}) = \{w'_{uv_j}w''_{uv_j}\} \cup A_{uv_j}^2$, and
    \item $\vec{F} \cap A_{uv_k} = \{w'_{uv_k}w''_{uv_k},w''_{uv_k}w'_{uv_k}\} = \{w'_{uv_k}w''_{uv_k}\} \cup A_{uv_k}^3$.
\end{itemize}

Now let $F_\ell \subseteq E(H)$ contain all edges $e$ such that $\vec{F} \cap A_{e} = \{w'_{e}w''_{e}\} \cup A_{e}^\ell$. It follows directly from the above observation that each $F_\ell$ is a perfect matching in $H$.
Thus, we can obtain a proper edge-coloring $\phi$ of $H$ by $\phi(e) \coloneqq \ell$ for each $e \in F_\ell$.
Furthermore, by construction of $D$, if $\phi(e) = \ell$ then $\ell \in \Phi(e)$.
This means that $\phi$ is a list edge-coloring of $(H, \Phi)$, which completes the proof.
\end{proof}

Thus, we complete the proof of Lemma~\ref{lem:planar_even_odd}.
\end{proof}

\begin{remark}\label{rem:undirected_planar_cubic}
Note that \textsc{$\forall$Even 2-Factor} for a cubic undirected graph $H$ is essentially the same as finding a $3$-edge-coloring in the same graph $H$.
Indeed, as in the proof of Claim~\ref{vgudaiosian}, the equivalence is observed as follows:
\begin{itemize}
\item a $3$-edge-coloring $\phi$ corresponds to a partition of $E(H)$ into three perfect matchings, two of which form a $2$-factor $F$ in which all the cycles are even;
\item for a $2$-factor $F$ in $H$, $E(H) \setminus F$ is a perfect matching, and if all the cycles in $F$ are even, then $F$ can be decomposed into two perfect matchings.
\end{itemize}
Finding a $3$-edge-coloring for cubic graphs is known to be $\NP$-hard~\cite{holyer1981NPedge}, but it can be done in polynomial time for planar cubic graphs and the complexity for planar graphs is open even when the graph is subcubic; see~\cite[Section~7]{goetze2026recognition} for more details.
\end{remark}

\section{Undirected Problems}
\label{sec:undirected}
Our goal in this section is to prove the following theorem.

\begin{theorem}\label{thm:undirected}
    \textsc{$\forall$Odd $2$-Factor}, \textsc{$\forall$Even $2$-Factor}, \textsc{$\exists$Odd $2$-Factor}, and \textsc{$\exists$Even $2$-Factor} are $\NP$-complete even for subcubic undirected graphs.
\end{theorem}

To prove this theorem, we show a transformation of a directed graph to an undirected graph with a natural bijection between the $(1, 1)$-factors to the $2$-factors which preserves the numbers of even cycles and of odd cycles.
Given an undirected graph $G$ and $a,b \in \mathbb{Z}_{\geq 0}$, an \emph{$(a,b,2)$-factor} is a $2$-factor of $G$ that consists of $a$ even cycles and $b$ odd cycles. Similarly, given a directed graph $D$ and $a,b \in \mathbb{Z}_{\geq 0}$, an \emph{$(a,b,1,1)$-factor} is a $(1,1)$-factor of $D$ that consists of $a$ even directed cycles and $b$ odd directed cycles.

The main lemma here is stated as Lemma~\ref{lem:connection_digraph_undirected}.
Together with Theorems~\ref{thm:directed} and~\ref{thm:directed_planar}, it immediately implies Theorem~\ref{thm:undirected} and Corollary~\ref{cor:undirected_planar}, respectively.

\begin{corollary}\label{cor:undirected_planar}
\textsc{$\forall$Odd $2$-Factor} is $\NP$-hard even for planar subcubic graphs.
\end{corollary} 

\begin{lemma}\label{lem:connection_digraph_undirected}
    Given a digraph $D$ such that $\deg_D^-(v) \ge 1$ and $\deg_D^+(v) \ge 1$ for every vertex $v$, we can compute in polynomial time an undirected graph $G$ such that for any  $a,b \in \mathbb{Z}_{\geq 0}$, we have that $G$ contains an $(a,b,2)$-factor if and only if $D$ contains an $(a,b,1,1)$-factor.
    The maximum degree of a vertex in $G$ is the maximum of the in-degree and out-degree of a vertex in $D$ plus 1.
    Moreover, if $D$ is subcubic and planar, then $G$ is also subcubic and planar.
\end{lemma}

\begin{proof}
Let $D$ be a digraph. We construct a graph $G$ in the following way. We let $V(G)$ consist of three vertices $v^-, v', v^+$ for all $v \in V(D)$. Next, for all $v \in V(D)$, we let $E(G)$ contain two edges $v^-v'$ and $v'v^+$. Finally, for every $uv \in A(D)$, we let $E(G)$ contain the edge $u^+v^-$. This completes the description of $G$. It is easy to see that $G$ can be built from $D$ in polynomial time, and it satisfies the stated maximum degree condition.
Moreover, if $D$ is subcubic, then each vertex is either of in-degree $1$ or of out-degree $1$, so it is easy to preserve the planarity (around each $v \in V(D)$, the incoming arcs and the outgoing arcs are respectively incident in a consecutive way, so one can easily arrange $v^-, v', v^+ \in V(G)$ around $v$ accordingly).
An illustration appears in Figure~\ref{fig:construction}.

\begin{figure}[t!]
    \centering
    \begin{subfigure}{.3\textwidth}    
    \centering
        \begin{tikzpicture}[
            myNode/.style={circle, draw,  minimum size=2em},
            myArrow/.style={draw,line width = 1pt, -{Stealth[length=3mm]}}]
        
            \node (u) at (0,0) [myNode] {$u$};
            \node (v) at (1.5,1.5) [myNode] {$v$};
            \node (w) at (3,0) [myNode] {$w$};
        
            \foreach \u\v in {u/v,v/w,w/u}{
                \draw[myArrow] (\u) to (\v); 
            }
        \end{tikzpicture}
        \caption{Example of digraph $D$.}
        \label{fig:construction-digraph}
    \end{subfigure}\hfill
    \begin{subfigure}{.65\textwidth}
    \centering
    \begin{tikzpicture}[
        myNode/.style={circle, draw,  minimum size=2em},
        myEdge/.style={draw,line width = 1pt]}]
    
        \node (bu) at (0,-1.5) [myNode] {$u'$};
        \node (cu) [myNode, above left = 1em and 1em of bu] {$u^+$};
        \node (au) [myNode, above right = 1em and 1em of bu] {$u^-$};
        \node (bv) at (2,1.5) [myNode] {$v'$};
        \node (av) [myNode, below left = 1em and 1em of bv] {$v^-$};
        \node (cv) [myNode, below right = 1em and 1em of bv] {$v^+$};
        \node (bw) at (4,-1.5) [myNode] {$w'$};
        \node (cw) [myNode, above left = 1em and 1em of bw] {$w^+$};
        \node (aw) [myNode, above right = 1em and 1em of bw] {$w^-$};
    
        \foreach \u\v in {cw/au,cu/bu,bu/au,av/bv,bv/cv,cw/bw,bw/aw}{
            \draw[myEdge] (\u) to (\v); 
        }
        \draw[myEdge] (cu) to (av);
        \draw[myEdge] (cv) to (aw);
    \end{tikzpicture}
        \caption{Final construction of undirected graph $G$ from $D$.}
        \label{fig:construction-undirected}
    \end{subfigure}
    \caption{Illustration for Lemma~\ref{lem:connection_digraph_undirected}.}
    \label{fig:construction}
\end{figure}

The following claim completes the proof, which is intuitively easy to see (as in the end of Section~\ref{sec:proof_of_reduction}).

\begin{claim}\label{cl:undirected}
    $D$ contains an $(a,b,1,1)$-factor if and only if $G$ contains an $(a,b,2)$-factor.
\end{claim}

\begin{proof}
First suppose that $D$ contains an $(a,b,1,1)$-factor $\vec{F}$. Let $F \subseteq E(G)$ be the set consisting of $v^-v'$ and $v'v^+$ for all $v \in V(D)$ and the edges $u^+v^-$ for all $u,v \in V(D)$ with $uv \in \vec{F}$.
We will show that $F$ is an $(a,b,2)$-factor of $G$. First observe that for every $v \in V(D)$, we have $\deg_F(v^-)=\deg_{\vec{F}}^-(v)+1=2$, $\deg_F(v')=2$, and $\deg_F(v^+)=\deg_{\vec{F}}^+(v)+1=2$, so $F$ is a 2-factor of $G$. Next, we describe a bijection $f\colon\mathcal{C}(\vec{F})\rightarrow \mathcal{C}(F)$. Namely, let $\vec{C}\in \mathcal{C}(\vec{F})$. Then, we let $f(\vec{C})$ be defined by $E(f(\vec{C}))=\bigcup_{v \in V(\vec{C})}\{v^-v',v'v^+\}\cup \bigcup_{uv \in A(\vec{C})}\{u^+v^-\}$. It is not difficult to see that $f$ is a bijection from $\mathcal{C}(\vec{F})$ to $\mathcal{C}(F)$ indeed. Moreover, for every $\vec{C}\in \mathcal{C}(\vec{F})$, we have $|E(f(\vec{C}))|=3|A(\vec{C})|$. In particular, we obtain that $|E(f(\vec{C}))|$ and $|A(\vec{C})|$ have the same parity. It follows that $F$ is an $(a,b,2)$-factor of $G$.

Now suppose that $G$ contains an $(a,b,2)$-factor $F$. We define a set $\vec{F}\subseteq A(D)$. Namely, we let $\vec{F}$ contain the arcs $uv \in A(D)$ for all pairs of $u,v \in V(D)$ with $u^+v^-\in F$. Observe that $\vec{F}$ is well-defined by the construction of $G$. We show that $\vec{F}$ is an $(a,b,1,1)$-factor of $D$. Note that for every $v\in V(D)$, as $F$ is a 2-factor of $G$ and by the construction of $G$, we have that $\{v^-v', v'v^+\}\subseteq F$. As $F$ is a 2-factor of $G$, for every $v \in V(D)$, there exists exactly one edge in $(\{ v^-u \mid u\in V(G) \}\setminus \{v^-v'\})\cap F$. It follows by construction that this edge is of the form $u^+v^-$ for some $u \in V(D)$ with $uv \in A(D)$. By the definition of $\vec{F}$, we get that $\deg_{\vec{F}}^-(v)=1$. A similar argument shows $\deg_{\vec{F}}^+(v)=1$, so $\vec{F}$ is a $(1,1)$-factor of $D$. Next, we describe a bijection $g\colon\mathcal{C}(F)\rightarrow \mathcal{C}(\vec{F})$. Namely, let $C\in \mathcal{C}(F)$. Then, we let $g(C)$ be defined by $A(g(C))=\{uv \in A(D) \mid u^+v^-\in E(C)\}$. It is easy to see that $g$ is a bijection from $\mathcal{C}(F)$ to $\mathcal{C}(\vec{F})$ indeed. Also, for every $C\in \mathcal{C}(F)$, we have $3|A(g(C))|=|E(C)|$. In particular, we get that $|A(g(C))|$ and $|E(C)|$ have the same parity. Hence $\vec{F}$ is an $(a,b,1,1)$-factor of $D$.
\end{proof}

Thus, we complete the proof of Lemma~\ref{lem:connection_digraph_undirected}.
\end{proof}

\section{Mixed Problem}\label{sec:mixed}
A \emph{mixed graph} is a graph that can contain both undirected edges and directed arcs.
For a mixed graph $G$, let $E(G)$ and $A(G)$ be the sets of edges and of arcs in $G$, respectively.
A path and a cycle are defined analogously:
$P$ is an \emph{$(s,t)$-path} if each edge in $E(G)$ can be oriented so that $P$ is a dipath in the resulting digraph.
Then, a \emph{cycle-factor} is also analogously defined.

In contrast to the undirected and directed cases, the problem \textsc{Cycle-Factor} in mixed graphs is already $\NP$-hard without any additional constraint.
This fact can be derived from the hardness result of Bang-Jensen and Casselgren~\cite{bang2015restricted}, as we show below in a stronger form.

Note that the naive approach of replacing each edge with two opposite arcs does not produce an equivalent instance of \textsc{$(1,1)$-Factor}.
It allows to use both arcs of a digon originated by the same edge $e \in E(G)$, i.e., to traverse $e$ in both directions as a $2$-cycle.
This does not correspond to any use of that edge in a cycle-factor of the original mixed graph (unless there are parallel edges).

In fact, \cite[Proposition 1.7]{bang2015restricted} shows that the problem of testing the existence of a $(1, 1)$-factor $F$ in a bipartite digraph such that $F$ contains no digon is $\NP$-hard.
This problem indeed corresponds to \textsc{Cycle-Factor} in mixed graphs: a digraph $D$ has such a $(1, 1)$-factor if and only if the mixed graph $H$ obtained from $D$ by replacing each digon with an edge has a cycle-factor.
By refining the reduction, we can obtain the following stronger result, where a mixed graph is called \emph{subcubic} if every vertex is incident to at most three edges and arcs in total.

\begin{theorem}\label{thm:mixed}
    \textsc{Cycle-Factor} in mixed graphs is $\NP$-hard even for subcubic bipartite graphs.
\end{theorem}

\begin{proof}
We reduce the following $\NP$-hard problem~\cite[Problem~2]{casselgren2012avoiding}.
\problemdef{Partial $3$-Edge-Coloring Extension}
    {A cubic bipartite undirected graph $H$ and a partial $3$-edge-coloring $\phi_X \colon X \to [3]$ with $X \subseteq E(H)$ such that $\phi_X(e) \in [2]$ for every $e \in X$.}
    {Find a $3$-edge-coloring $\phi \colon E(H) \to [3]$ with $\phi(e) = \phi_X(e)$ for every $e \in X$.}
    
Let $(V^+, V^-)$ be the bipartition of the vertex set $V(H)$; as $H$ is regular, we have $|V^+| = |V^-|$.
Let $D'$ be the digraph obtained from $H$ by replacing each edge with a digon, i.e., $A(D') = \{\, uv, vu \mid uv \in E(H) \,\}$.
We first construct a digraph $D$ on $V(H)$ from $D'$ by removing arcs according to $\phi_X$, and then transform $D$ into a subcubic mixed graph $G$ by replacing each remaining digon with an edge.

We define the set $R \subseteq A(D')$ of removed arcs as follows.
For each edge $e = uv \in X$ with $u \in V^+$ and $v \in V^-$,
\begin{itemize}
    \item if $\phi_X(e) = 1$, then $R$ contains all outgoing arcs of $u$ and all incoming arcs of $v$ except for the arc $uv$;
    \item if $\phi_X(e) = 2$, then $R$ contains all incoming arcs of $u$ and all outgoing arcs of $v$ except for the arc $vu$.
\end{itemize}
Define $D$ by $V(D) \coloneqq V(D') = V(H)$ and $A(D) \coloneqq A(D') \setminus R$.
Since $\phi_X$ is a proper edge-coloring of $X$ (otherwise, the answer to the original problem is clearly No), we have $\deg_D^-(v) \ge 1$ and $\deg_D^+(v) \ge 1$ for every $v \in V(D)$.

Let $G$ be the mixed graph obtained from $D$ by replacing each digon with an edge.
Then, $G$ is clearly subcubic and bipartite, and the next claim completes the proof.

\begin{claim}
    $H$ has a $3$-edge-coloring $\phi$ extending $\phi_X$ if and only if $G$ has a cycle-factor.
\end{claim}

\begin{proof}
Suppose that there exists a $3$-edge-coloring $\phi \colon E(H) \to [3]$ with $\phi(e) = \phi_X(e)$ for every $e \in X$.
We define two arc sets in $D'$ as follows:
\begin{align*}
    \vec{F}_1 \coloneqq \{\, uv \mid e = uv \in E(H),~u \in V^+,~v \in V^-,~\phi(e) = 1 \,\},\\
    \vec{F}_2 \coloneqq \{\, vu \mid e = uv \in E(H),~u \in V^+,~v \in V^-,~\phi(e) = 2 \,\}.
\end{align*}
Since $\phi$ is a $3$-edge-coloring of a cubic graph, $\vec{F} \coloneqq \vec{F}_1 \cup \vec{F}_2$ is a $(1, 1)$-factor of $D'$.
Then, to see the existence of a cycle-factor $F$ of $G$ (corresponding to $\vec{F}$), it suffices to show that for every arc $uv \in \vec{F}$ there exists either an arc $uv \in A(G)$ or an edge $uv \in E(G)$.
By construction, they are rephrased as either $uv \in A(D)$ and $vu \not\in A(D)$ or $uv \in A(D)$ and $vu \in A(D)$, and hence just $uv \in A(D)$, equivalently, $uv \not\in R$.

Let $\vec{e} = uv \in \vec{F}$.
We show $\vec{e} \not\in R$.
If the underlying edge $e = uv \in E(H)$ belongs to $X$, then $\vec{e} \not\in R$ by construction.
Suppose that $e \not\in X$.
If $\vec{e} \in \vec{F}_1$, then $u \in V^+$ and $v \in V^-$.
In this case, if $\vec{e} \in R$, then there exists an edge $e' \neq e$ in the form of $uv'$ or $u'v$ with $\phi(e') = \phi_X(e') = 1$ by construction, but this contradicts $\phi(e) = 1$.
Thus, $\vec{e} \not\in R$.
The case where $\vec{e} \in \vec{F}_2$ is similar.
Therefore, we conclude that $\vec{e} \not\in R$ in either case, which implies that there exists a cycle-factor $F$ of $G$ that has an orientation $\vec{F}$.

Suppose that there exists a cycle-factor $F$ of $G$.
Let $\vec{F} \subseteq A(D)$ be a $(1, 1)$-factor of $D$ obtained as an orientation of $F$.
We partition it into two arc sets in $D'$ as follows:
\begin{align*}
    \vec{F}_1 \coloneqq \{\, uv \mid uv \in \vec{F},~u \in V^+,~v \in V^- \,\},\\
    \vec{F}_2 \coloneqq \{\, vu \mid vu \in \vec{F},~u \in V^+,~v \in V^- \,\}.
\end{align*}
For each $i \in [2]$, let $F_i \subseteq E(H)$ be the set of corresponding edges in $H$, and let $F_3 \coloneqq E(H) \setminus (F_1 \cup F_2)$.
Since $\vec{F}$ is a $(1, 1)$-factor of $D'$ and $H$ is a cubic bipartite graph, each $F_i$ $(i \in [3])$ is a perfect matching.
Thus, one can obtain a $3$-edge-coloring $\phi$ of $H$ by $\phi(e) \coloneqq i$ $(e \in F_i)$ for each $i \in [3]$.

We finally confirm that $\phi(e) = \phi_X(e)$ for every $e \in X$.
Suppose to the contrary that $\phi(e) \neq \phi_X(e)$ for some $e = uv \in X$.
By symmetry, we may assume that $\phi_X(e) = 1$ and $\phi(e) \neq 1$.
By definition of $R$, the arc $uv$ is a unique outgoing arc for $u$ and a unique incoming arc for $v$ in $D$.
Thus, $uv$ must be included in any $(1, 1)$-factor of $D$, including $\vec{F}$.
This implies that $uv \in \vec{F}_1$, which contradicts $\phi(e) \neq 1$.
Thus, we conclude that $\phi(e) = \phi_X(e)$ for every $e \in X$, which completes the proof.
\end{proof}

Thus, we complete the proof of Theorem~\ref{thm:mixed}.
\end{proof}

\subsection*{Acknowledgments}
We would like to thank Kristóf Bérczi for initial discussions on the problem and organization of the 16th and 17th Emléktábla Workshops in July 2024 and July 2025, respectively, where the collaboration of the authors was hosted and supported.
We also thank J{\o}rgen Bang-Jensen for pointing out important previous work.
We are grateful to Kunihiro Wasa for his AI use to solve the remaining open problem in the earlier version~\cite{horsch2025odd} of this paper, which is detailed in the next section.

Csaba Kir\'aly was supported by the Hungarian Scientific Research Fund (OTKA) grant PD138102 with the support provided from the National Research, Development and Innovation Fund of Hungary, financed under the PD\_21 funding scheme. 
Mirabel Mendoza-Cadena was supported by Centro de Modelamiento Matemático (CMM) BASAL fund FB210005 for center of excellence from ANID-Chile. 
Gyula Pap was supported by the Hungarian National Research, Development and Innovation Office grant NKFI-132524.
Yutaro Yamaguchi was supported by JSPS KAKENHI Grant Numbers 20K19743, 20H00605, and 25H01114, by JST CRONOS Japan Grant Number JPMJCS24K2, and by JST ASPIRE Japan Grant Number JPMJAP2520. 

\subsection*{Tools and Computational Resource Disclosure}
\textsc{$\exists$Even Cycle-Factor} had been the only remaining problem in the earlier version~\cite{horsch2025odd} of this paper, and ChatGPT 5.6 Pro used by Kunihiro Wasa returned a prototype of its $\NP$-hardness proof with the key reference~\cite{BARRINGTON1989150}.
The unified proof presented in this paper that covers all the four variants for subcubic graphs has been simplified by the authors, but this was a significant step. 
During the preparation of this manuscript, the authors used ChatGPT (OpenAI) and Claude (Anthropic) to assist with editing, grammar, and formatting, and with verifying and streamlining proofs.
The authors verified and reviewed all AI-assisted edits and take full responsibility for all content within this manuscript.

\bibliographystyle{plain}
\bibliography{factorCycle_v2}
\end{document}